\magnification=\magstep1
\tolerance 500
\rightline{TAUP 2588-99}
\rightline{5 October 1999}
\vskip  2 true cm
\centerline{\bf Representation of the Resonances}
\centerline{\bf of a } 
\centerline{\bf Relativistic Quantum Field Theoretical Model}
\centerline{\bf in}
\centerline{\bf Lax-Phillips Scattering Theory}
\smallskip
\centerline{Y. Strauss and L.P. Horwitz\footnote{*}{Also at Department of
Physics, Bar Ilan University, Ramat Gan 52900, Israel.   e-mail:larry@
post.tau.ac.il}}
\centerline{School of Physics}
\centerline{Raymond and Beverly Sackler Faculty of Exact Sciences}
\centerline{Tel Aviv University}
\centerline{Ramat Aviv 69978, Israel}
\bigskip
{\it Abstract:\/}   We apply the quantum Lax-Phillips scattering theory
 to a relativistically covariant quantum field theoretical
form of the (soluble) Lee model.  We construct the translation
representations with the help of the wave operators, and show that the
resulting Lax-Phillips $S$-matrix is an inner function (the
Lax-Phillips theory is essentially a theory of translation invariant
subspaces). We then discuss the non-relativistic limit of this theory,
 and show that the resulting kinematic relations coincide with the
 conditions required for the Galilean description of a decaying system.   
\vfill
\eject
\noindent
{\bf 1. Introduction.}
\par The theory of Lax and Phillips$^1$ (1967), originally developed for the 
description of resonances in electromagnetic or acoustic scattering phenomena,
has been  used as a framework for the construction of a description of
 irreversible resonant phenomena in the
 quantum theory$^{2-5}$ (which we will refer to as the quantum Lax-Phillips 
theory). This leads to a time evolution of resonant states
which is of semigroup type, i.e., essentially exponential decay.  Semigroup
evolution is necessarily a property of irreversible processes$^6$. It 
appears experimentally that elementary particle decay, to a high 
degree of accuracy, follows a semigroup law, and hence such processes seem 
to be irreversible.    
\par The theory of Weisskopf and Wigner$^7$, which 
is based on the definition of the survival amplitude of the initial state
$\phi$ (associated with the unstable system) as the scalar 
product of that state
with the unitarily evolved state, 
$$  (\phi, e^{-iHt}\phi) \eqno(1.1)$$
cannot have exact exponential behavior$^8$. One can easily generalize this 
construction to the problem of more than one resonance$^{9,10}$. If $P$ is the 
projection operator into the  subspace of initial states
($N$-dimensional for $N$ resonances),
the reduced evolution operator is given by
$$Pe^{-iHt} P .\eqno(1.1')$$
 This operator cannot be an element of a semigroup.$^8$
\par  Experiments 
on the decay of 
the neutral $K$-meson system$^{11}$ show clearly that the 
phenomenological description of Lee, Oehme and Yang$^{12}$, and Wu and 
Yang$^{13}$, by means of a $2\times 2$ effective Hamiltonian which corresponds
to an exact semigroup evolution of the unstable system, provides a very
accurate description of the data. Since the Wigner-Weisskopf
theory cannot provide a semigroup evolution law$^8$, the
effective $2\times 2$ Hamiltonian cannot emerge in the framework of this
theory. Furthermore, it has been shown, using estimates based on
the quantum mechanical Lee-Friedrichs model$^{14}$, that the experimental
results appear to rule out the application of the Wigner-Weisskopf theory to
the decay of the neutral $K$-meson system.
\par  While an exponential decay
law can be derived explicitly in terms of a Gel'fand triple$^{15}$, the 
representation of the resonant state in this framework is in a Banach space
which does not, in general,  coincide with a quantum mechanical 
Hilbert space; it 
 does not have the properties of a Hilbert space, such as scalar products 
and the possibility of calculating expectation values.
One cannot compute physical properties other than the lifetimes in this way.
\par The quantum Lax-Phillips 
theory provides the possibility of constructing a fundamental theoretical 
description of the resonant system which has exact semigroup evolution,
and represents the resonance as a {\it state in a Hilbert space}. The
Lax-Phillips theory, based on certain classes of functions with
half-line support properties, necessarily deals with families of
Hilbert spaces of
Hardy class functions. We conjecture that the Gel'fand triples
constructed on such spaces for the description of
resonant states can be imbedded in the large Hilbert
space of the Lax-Phillips theory. 
\par Progess has recently been made on the application of stochastic
 methods$^{16}$ to construct a generalization of the Schr\"odinger
 evolution which describes collapse of the wave function during the
 measurement process, and may indeed provide a framework for general
 irreversible processes, such as the particle decay problem with which
 we shall be concerned here.  The work of Parthsarathy and Hudson$^{17}$,
 imbedding such processes in a Hilbert space (in which martingales may
 be represented as semigroups) leads us to conjecture a
 close relation with the Lax-Phillips theory. 
\par These conjectures, if realized, would establish the Lax Phillips
 theory as a unifying framework for several approaches to the
 description of irreversible processes. We confine ouselves in
 the present work to a discussion of the quantum Lax-Phillips theory.
\par In the 
following, we describe briefly the structure of the quantum
Lax-Phillips theory, and give
some physical interpretation for the states of the Lax-Phillips Hilbert space. 
\par The Lax-Phillips theory is defined in a Hilbert space $\overline{\cal H}$
 of states which 
contains two distinguished subspaces, ${\cal D}_\pm$, called ``outgoing'' and 
``incoming''.  There is a unitary evolution law which we denote
 by $U(\tau)$, for which these subspaces are invariant in the following
sense:
$$\eqalign{ U(\tau) {\cal D}_+ &\subset {\cal D}_+ \qquad  \tau \geq 0 \cr
U(\tau) {\cal D}_- &\subset {\cal D}_- \qquad \tau \leq 0 \cr} \eqno(1.2)$$
\par The translates of ${\cal D}_\pm$ under $U(\tau)$ are dense, i.e.,
$$ {\overline {{\bigcup_\tau}\, U(\tau) {\cal D}_\pm}} =  {\overline {\cal H}}
  \eqno(1.3)$$
and the asymptotic property
$$ {\bigcap_\tau }\,U(\tau) {\cal D}_\pm = \emptyset \eqno(1.4)$$
is assumed. It follows from these properties that 
$$ Z(\tau) = P_+ U(\tau) P_-, \eqno(1.5)$$
where $P_\pm$ are projections into the subspaces orthogonal to ${\cal D}_\pm$,
is a strongly contractive semigroup$^1$, i.e.,
$$ Z(\tau_1) Z(\tau_2) = Z(\tau_1 + \tau_2) \eqno(1.6)$$
for $\tau_1,\, \tau_2$ positive, and $\Vert Z(\tau)\Vert \to 0$ for $\tau \to
0$. It follows from $(1.2)$ that $Z(\tau)$ takes the subspace
$\cal K$, the orthogonal complement of ${\cal D}_\pm$ in $\overline{\cal H}$
(associated with the resonances in the Lax-Phillips theory),
into itself$^1$, {\it i.e.},
$$ Z(\tau)= P_{\cal K} U(\tau) P_{\cal K}. \eqno(1.7)$$
The relation $(1.7)$ is of the same structure as $(1.1')$; there is, as we 
shall see in the following, an essential difference in the way that the
subspaces associated with resonances are defined. The argument that $(1.1')$
cannot form a semigroup is not valid$^3$ for $(1.7)$. 
\par There is a theorem of Sinai$^{18}$ which affirms that a Hilbert 
space with the properties that there are distinguished subspaces 
 satisfying,
with a given law of evolution  $U(\tau)$, the properties $(1.2),\,(1.3),
\, (1.4)$ has a foliation into a one-parameter (which we shall denote as $s$)
family of isomorphic Hilbert spaces, which are called auxiliary
Hilbert spaces, ${\cal H}_s$ for which
$$ \overline{\cal H} = {\int_\oplus} {\cal H}_s. \eqno(1.8)$$
Representing these spaces in terms of square-integrable functions, we define
the norm in the direct integral space (we use Lebesgue measure)
as 
$$ \Vert f \Vert^2 = \int_{-\infty}^\infty ds \Vert f_s\Vert^2_H, \eqno(1.9)$$
where $f \in {\overline H} $ represents a vector in ${\overline{\cal H}}$
 in terms of a function in the
$L^2$ function space ${\overline H}=L^2(-\infty, \infty, H)$;  $f_s$ is 
an element of $H$, the $L^2$ 
function space  (which we shall call the {\it auxiliary space})
representing ${\cal H}_s$  for any $s$ [we shall not add in what follows a
subscript to the norm or scalar product symbols for scalar products of elements
of the auxiliary Hilbert space associated to a point $s$ on the foliation
axis].
\par The Sinai theorem furthermore asserts that there are representations
for which the action of the full evolution group $U(\tau)$ on 
$L^2(-\infty, \infty;H)$ is translation by $\tau$ units. Given $D_\pm$ (the
subspaces of $L^2$ functions representing ${\cal D}_\pm$), there is such a
representation,
called the {\it incoming representation}$^1$, for which the set of all
functions in $D_-$ have support in $(-\infty, 0)$ and constitute the subspace
 $L^2(-\infty,0;H)$ of $L^2(-\infty, \infty;H)$; there is another
 representation,
called the {\it outgoing representation}, for which functions in $D_+$
have support in $(0,\infty)$ and 
constitute the subspace $L^2(0,\infty;H)$ of $L^2(-\infty, \infty;H)$.
The fact that $Z(\tau)$ in Eq. (1.7) is a semigroup is a consequence of
the definition of the subspaces $D_\pm$ in terms of support properties
on intervals along the foliation axis in the {\it outgoing} and {\it incoming}
translation representations respectively. The non self-adjoint character of
the generator of the semigroup $Z(\tau)$ is a consequence of this structure.
\par Lax and Phillips$^1$ show that there are unitary operators $W_\pm$, 
called wave operators, which map elements in ${\overline{\cal H}}$, 
respectively, to these representations.  They define an $S$-matrix, 
$$ S= W_+W_-^{-1}  \eqno(1.10)$$
which connects the incoming to the outgoing representations; it is
unitary, commutes with 
translations, and maps $L^2(-\infty,0;H)$ into itself (invariance of 
a subspace of Hardy class functions). Since $S$
commutes with translations, it is diagonal in Fourier (spectral)
representation.  As pointed out by
Lax and Phillips$^1$, according to a special case of a theorem of
Four\`es and Segal$^{19}$, an operator with these properties can be
represented
as a multiplicative operator-valued function ${\cal S}(\sigma)$ which
maps $H$ into $H$, and satisfies the following conditions:
$$\eqalign{ (a)\   &{\cal S}(\sigma)\  is\  the\  boundary\  
value\  of\  an\ \cr
&operator{\rm -}valued\  function\ {\cal S}(z)\ analytic\ 
for\ {\rm Im}z >0. \cr
(b)\   &\Vert {\cal S}(z) \Vert  \leq 1\ for\ all\ z\ with\  {\rm Im}z >0. \cr
(c)\   &{\cal S}(\sigma)\ is\ unitary\ for\ almost\ all\ real\ \sigma.\cr}$$ 
 An operator with
these properties is known as an inner function$^{20}$; such operators
arise in the study of shift invariant subspaces, the
essential mathematical content of the Lax-Phillips theory. 
 The singularities of
 this $S$-matrix, in what we shall define as the 
{\it spectral representation} (defined in terms of the Fourier
transform on the foliation variable $s$), coincide with the spectrum
of the generator of the semigroup characterizing the evolution 
of the unstable system.    
\par  In the framework of quantum theory, one may identify the Hilbert
space ${\cal H}$ with a space of physical states, and the variable
$\tau$ with the laboratory time (the semigroup
evolution is observed in the laboratory according to this time)$^2$.
  The representation
of this space in terms of the foliated $L^2$ space ${\overline H}$
provides a natural probabilistic interpretation for the auxiliary spaces 
associated with each value of the foliation variable $s$, i.e., the 
quantity  $\Vert f_s \Vert^2 $ corresponds to the probability density
for the system to be found in the neighborhood of $s$. For example,
consider an operator $A$ defined on ${\overline H}$ which acts pointwise, i.e.,
contains no shift along the foliation. Such an operator can be
represented as a direct integral
$$ A = \int_\oplus A_s.  \eqno(1.11)$$
It produces a map
of the auxiliary space $H$ into $H$ for each value of $s$, and thus,
if it is self-adjoint, $A_s$  may act as an observable in a quantum theory
associated to the point $s$;$^{4}$ The expectation value of $A_s$ in a state
in this Hilbert space defined by the vector $\psi_s$, the component of
$\psi \in {\overline H}$ in the auxiliary space at $s$, is 
$$ \langle A_s \rangle_s = {(\psi_s, A_s \psi_s) \over \Vert \psi_s
\Vert^2}
 \eqno(1.12) $$.
Taking into account the {\it a priori} probability density $\Vert
\psi_s\Vert^2$ that the system is found at this point on the foliation
axis, we see that the expectation value of $A$ in ${\overline H}$ is
$$ \langle A \rangle =  \int ds \langle A_s \rangle_s \Vert \psi_s\Vert^2
= \int ds (\psi_s, A_s \psi_s ), \eqno(1.13)$$    
the direct integral representation of $(\psi, A \psi)$.
\par As we have remarked above, in the translation representations for
 $U(\tau)$ the foliation variable $s$ is shifted (this shift, for
 sufficiently large $\vert \tau \vert$,  induces the transition of
 the state into the subspaces ${\cal D}_\pm$). It follows that $s$
 may be identified as an intrinsic time associated with the evolution
 of the state; since it is a variable of the measure space of the
 Hilbert space ${\overline{\cal H}}$, this quantity itself has the
 meaning of a quantum variable.   
\par We are presented here with the notion of a virtual history. To
understand this idea, suppose that at a given time $\tau_0$, the
function which represents the state has some distribution $\Vert
\psi_s^{\tau_0}\Vert^2$. This distribution provides an {\it a priori}
probability that the system would be found at time $s$ (not
 necessarily equal to 
$\tau_0$), if the experiment were performed at time $s$ corresponding 
 to $\tau = s$ on the laboratory clock.  The state of the system
therefore contains information on the structure of the {\it history} of
 the system as it is inferred at $\tau_0$.   
\par We shall assume the existence of a unitary evolution on the
Hilbert space $\overline{{\cal H}}$, and that for
$$ U(\tau) = e^{-iK\tau}, \eqno(1.14)$$ 
the generator $K$ can be decomposed as 
$$ K= K_0 + V \eqno(1.15)$$
 in terms of an unperturbed
operator $K_0$ with spectrum $(-\infty, \infty)$ and a perturbation
$V$, under which this spectrum is stable.  
We shall, furthermore, assume that wave operators exist, defined on
some dense set, as 
$$ \Omega_\pm = \lim_{\tau \rightarrow \pm \infty} e^{iK\tau}e^{-iK_0
\tau}. \eqno(1.16)$$
In the soluble model that we shall treat as an example here,,
 the existence of the wave operators is assured.
\par  With the help of the
 wave operators, we can define translation representations for $U(\tau)$. 
 The translation representation for $K_0$ is defined by the property
$$ {{}_0\langle} s, \alpha \vert e^{-iK_0\tau} f)= {{}_0\langle} s-\tau, \alpha
\vert f), \eqno(1.17)$$
where $\alpha$ corresponds to a label for the basis of the auxiliary space.  
Noting that
$$  K \Omega_\pm = \Omega_\pm K_0 \eqno(1.18) $$
we see that
$$ {{}_{out \atop in}\langle} s, \alpha \vert e^{-iK\tau} f) =
 {{}_{out \atop in} \langle}
s-\tau, \alpha \vert f),  \eqno(1.19)$$
where
$$ {{}_{out \atop in}\langle}s, \alpha \vert f) = {{}_0 \langle}s,
  \alpha \vert 
\Omega_\pm^\dagger f) \eqno(1.20)$$
\par It will be convenient to work in terms of the Fourier transform of
 the {\it in} and {\it out} 
translation representations; we shall call these  the {\it in} and
 {\it out} {\it spectral}
representations, {\it i.e.},
$$ {{}_{out \atop in}\langle} \sigma, \alpha \vert f) = \int_{-\infty}^\infty
e^{-\sigma s} {{}_{out \atop in}\langle} s, \alpha \vert f). \eqno(1.21)$$
In these representations, (1.20) is
$$ {{}_{out \atop in}\langle}\sigma, \alpha \vert f) = {{}_0 \langle}\sigma,
   \alpha \vert \Omega_\pm^\dagger f)
    \eqno(1.22)$$
and $(1.19)$ becomes
$$ {{}_{out \atop in}\langle} \sigma, \alpha \vert e^{-iK\tau}\vert f) =
  e^{-i\sigma \tau}{{}_{out \atop in}\langle} \sigma, \alpha \vert f).
\eqno(1.23)$$
Eq. $(1.17)$ becomes, under Fourier transform
$$ {{}_0\langle} \sigma, \alpha \vert e^{-iK_0\tau} f)= e^{-i\sigma \tau}
{{}_0\langle} \sigma, \alpha \vert f). \eqno(1.24)$$
For $f$ in the domain of $K_0$, $(1.23)$ implies that
$$ {{}_0\langle} \sigma, \alpha \vert K_0 f)= \sigma
{{}_0\langle} \sigma, \alpha \vert f). \eqno(1.25)$$
\par With the solution of $(1.25)$, and the wave operators,
the {\it in} and {\it out} spectral 
representations of a vector $f$ can be constructed from $(1.24)$.
\par We are now in a position to construct the subspaces ${\cal
D}_\pm$, which are not given, {\it a priori}, in the Lax-Phillips quantum
theory.  Identifying ${{}_{out} \langle}s, \alpha \vert f)$ with the  {\it
outgoing} translation representation, we shall define $ D_+$ as the
set of functions with  support in $(0,\infty)$ in this
representation. Similarly, identifying ${{}_{in} \langle}s, \alpha \vert
f)$ with the  {\it
incoming} translation representation, we shall define $ D_-$ as the
set of functions with  support in $(-\infty,0)$ in this
representation. The corresponding elements of ${\overline{\cal H}}$
constitute the subspaces ${\cal D}_\pm$.   By construction,
${\cal  D}_\pm$ have the required invariance properties under the
action of $U(\tau)$.
\par  The {\it outgoing spectral representation}
of a vector $g \in {\cal H}$ is
$$\eqalign{  {{}_{out}\langle} \sigma \alpha \vert g ) =
{{}_0\langle}\sigma\alpha \vert \Omega_+^{-1} g)&= 
\int\,d\sigma' \sum_{\alpha'} {{}_0 \langle} \sigma \alpha \vert {\bf
S}\vert \sigma' \alpha' \rangle_0 {{}_0 \langle}\sigma' \alpha' \vert
\Omega_-^{-1} g)\cr & = \int\,d\sigma' \sum_{\alpha'} {{}_0 \langle} 
\sigma \alpha \vert {\bf
S}\vert \sigma' \alpha' \rangle_0 {{}_{in} \langle}\sigma' \alpha' 
\vert g),\cr} \eqno (1.26)$$
where we call 
$${\bf S} = \Omega_+^{-1} \Omega_-. \eqno(1.27)$$
the quantum Lax-Phillips $S$-operator. We see that the kernel ${{}_0 \langle} 
\sigma \alpha \vert {\bf
S}\vert \sigma' \alpha' \rangle_0$ maps the incoming to outgoing
spectral representations.  Since $\bf S$ commutes with $K_0$, it
follows that 
$$ {{}_0 \langle} 
\sigma \alpha \vert {\bf
S}\vert \sigma' \alpha' \rangle_0 = \delta(\sigma - \sigma') S^{\alpha
\alpha'}. (\sigma) \eqno(1.28)$$
It follows from $(1.16)$ and $(1.22)$, in the standard way$^{21}$, that
$$ {{}_0\langle} \sigma \alpha \vert {\bf S} \vert \sigma' \alpha' 
{\rangle_0} =  \lim_{\epsilon \rightarrow 0}
\delta(\sigma-\sigma')\{\delta^{\alpha \alpha'} - 
2\pi i {{}_0\langle} \sigma \alpha \vert {\bf T}(\sigma + i\epsilon)
 \vert \sigma' \alpha' {\rangle_0} \}, \eqno(1.29)$$
where
$$ {\bf T}(z) = V + VG(z)V = V + VG_0(z) {\bf T}(z). \eqno(1.30)$$
We remark that, by this construction,
$S^{\alpha \alpha'}(\sigma)$ is {\it analytic in the upper half plane} in 
$\sigma$.
The Lax-Phillips $S$-matrix$^1$ is given by the inverse Fourier transform,
$$ S = \bigl\{{{}_0 \langle} 
s \alpha \vert {\bf
S}\vert s' \alpha' \rangle_0\bigr\};  \eqno(1.31)$$
this operator clearly commutes with translations. 
\par From $(1.29)$ it follows that the property $(a)$ above is true. Property 
$(c)$, unitarity for real $\sigma$, is equivalent to asymptotic
completeness, a property which is stronger than the existence of wave
operators. For the relativistic Lee model, which  we shall treat in
this paper, this condition is satisfied.   In the model that we shall 
study here, we shall see that there is a wide class of potentials $V$ for
which the operator $S(\sigma)$ satisfies the property $(b)$ specified above.
\par In the next section, we review briefly the structure of the 
relativistic Lee model$^{22}$, and construct explicitly the
Lax-Phillips spectral representations and $S$-matrix. Introducing
auxiliary space variables, we then characterize the properties of the
finite rank Lee model potential which assure that the $S$-matrix is an
inner function, {\it i.e.}, that property $(b)$ listed above  is satisfied.
\bigskip
\noindent
{\bf 2. Relativistic Lee-Friedrichs Model}
\smallskip
\par In this section, we define the relativistic Lee model$^{22}$ in terms of
bosonic quantum fields on spacetime ($x \equiv x^\mu$). These fields  evolve
 with an invariant
evolution parameter$^{23}$ $\tau$ (which we identify here with the evolution
 parameter of the Lax-Phillips theory discussed above);
 at equal $\tau$, they satisfy
the commutation relations (with $\psi^\dagger$ as the canonical
conjugate field to $\psi$; the fields $\psi$, which satisfy first order
 evolution equations
as for  nonrelativistic
 Schr\"odinger fields, are just annihilation operators)
$$ [\psi_\tau(x), \psi_\tau^\dagger(x')] =
\delta^4(x-x'). \eqno(2.1)$$
We remark that Antoniou, {\it et el}$^{24}$, have constructed a relativistic
Lee model of a somewhat different type; their field equation is
second order in the derivative with respect to the variable conjugate
to the mass.
\par In momentum space, for which
$$\psi_\tau(p) = {1 \over (2\pi)^2}\int d^4x e^{-ip_\mu x^\mu}
\psi_\tau(x), \eqno(2.2)$$
this relation becomes
$$ [\psi_\tau(p), \psi_\tau^\dagger(p') ] = \delta^4(p-p').\eqno(2.3)$$
The manifestly covariant spacetime structure of these fields is
admissible when $E,{\bf p}$ are not {\it a priori} constrained by a
 sharp mass-shell relation. In the mass-shell limit (for which the
variations in $m^2$ defined by $E^2 - {\bf p}^2$ are small), multiplying both
sides of $(2.3)$ by $\Delta E = \Delta m^2 / 2E$,
 one obtains the usual commutation
relations for on-shell fields,
 $$ [{\tilde \psi}_\tau ({\bf p}), {\tilde \psi}^\dagger({\bf p})] =
2E\delta^3({\bf p} - {\bf p}' ),  \eqno(2.4)$$
where $\tilde \psi ({\bf p}) = \sqrt{\Delta m^2} \psi(p)$.
The generator of evolution 
$$ K= K_0 + V \eqno(2.5)$$       
for which the Heisenberg picture fields  are
$$ \psi_\tau(p) = e^{iK\tau} \psi_0(p) e^{-iK\tau} \eqno(2.6)$$
is given, in this model, as (we write $p^2= p_\mu p^\mu$, $k^2 = k_\mu
k^\mu$ in the following) 
$$\eqalign{K_0 = \int\, d^4p\bigl\{ {p^2 \over 2M_V} b^\dagger(p) b(p) &+ {p^2
\over 2M_N} a^\dagger_N(p) a_N(p) \bigr\} \cr
& +\int \, d^4 k { k^2 \over 2M_\theta } a_\theta^\dagger(k)
a_\theta(k) \cr}\eqno(2.7)$$  
and
$$ V = \int d^4p \int d^4k (f(k) b^{\dagger}(p) a_N(p-k) a_\theta(k) +
   f^*(k) b(p) a^{\dagger}_N(p-k) a^{\dagger}_\theta(k),
   \eqno(2.8)$$
describing the process $V \leftrightarrow N + \theta $. 
The fields $b(p),\, a_N(p)$ and $a_\theta$ are annihilation operators
for the  $V,\, N$, and $\theta$ particles, respectively and $M_v,M_N$ and
$M_\theta$ are the mass parameters for the fields$^{22}$.
\par  The operators $$\eqalign{ Q_1 &= \int \,d^4p [b^\dagger(p)b(p) +
a_N^\dagger(p)a_N(p)]\cr Q_2 &= \int \, d^4p [a_N^\dagger(p)a_N(p) -
a_\theta^\dagger(p) a_\theta(p) ] \cr} \eqno(2.9)$$
are conserved, enabling us to decompose the Fock space to sectors. We
shall study the problem in the lowest sector $Q_1=1, Q_2 = 0$, for
which there is just one $V$ {\it or} one $N$ and one $\theta$. In this sector
the generator of evolution $K$ can be rewritten in the form
$$K=\int d^4p K^p = \int d^4p (K^p_0+V^p)$$
where
$$K^p_0= {{p^2} \over 2M_V}b^\dagger(p)b(p)+ \int d^4k
  \left( {{(p-k)^2}\over 2M_N}+{{k^2} \over 2M_\theta} \right) a^\dagger_N
  (p-k)a^\dagger_\theta(k)a_\theta(k)a_N(p-k)$$
and
$$ V^p = \int d^4k \left(f(k) b^{\dagger}(p) a_N(p-k) a_\theta(k) +
   f^*(k) b(p) a^{\dagger}_N(p-k) a^{\dagger}_\theta(k)\right) $$
In this form it is clear that both $K$ and $K_0$ have a direct
integral structure. This implies a similar structure for the wave operators
$\Omega_{\pm}$ and the possibility of defining restricted wave operators 
$\Omega^p_{\pm}$ for each value of $p$. From the expression for $K_0^p$
we see that $\vert V(p) \rangle =b^\dagger(p)\vert 0 \rangle$ can be regarded
as a discret eigenstate of $K_0^p$ and, therefore, is annihilated by
$\Omega^p_{\pm}$. This, in turn, implies that
$\Omega_{\pm}\vert V(p) \rangle=0$ for every $p$.
\par We give the explicit solution in the following$^{25}$. The $S$
matrix takes the simple form
$$ {}_0\langle \sigma' \alpha'\vert S \vert \sigma \alpha \rangle_0=
\delta(\sigma -\sigma')\bigl\{ 1-2\pi i \int d^4p
   { (\vert n \rangle_{ p,\sigma})^{\alpha'} 
  ( \vert n \rangle_{ p,\sigma})^{\alpha *}
   \over h( p,\sigma+i\epsilon)} \bigr\},   \eqno(2.9) $$
where
$$ h(p,z) = z-{{p^2}\over 2M_V}-\int d^4k { {\vert f(k)\vert^2} \over
   z- {{(p-k)^2}\over 2M_N} -{{k^2}\over 2M_\theta}}
   \eqno(2.10)$$
is the well-known denominator function of the Lee-Friedrichs model,
and
$$  (\vert n \rangle_{ p,\sigma})^{\alpha}= \int d^4k f^*(k) \langle
N(p-k) \theta(k) \vert \sigma \alpha
(N\theta)\rangle^*_0. \eqno(2.11)$$
On the pole (which occurs in the $N,\theta$ channel), the numerator
projects out the auxiliary space vector corresponding to the resonant
pole.  Note that the complex pole is in $\sigma \rightarrow z$, not in
$p^\mu$, so the energy momentum remains real.
We prove$^{25}$ (a theorem on Nevanlinna  class functions) that
$$ P_{n,P}(\sigma)\equiv
   {{\vert n \rangle_{\sigma,P} \langle n \vert_{\sigma,P}}
   \over {\scriptstyle \sigma,P}\langle n \vert n \rangle_{\sigma,P}}
   =\vert n \rangle_P{{}_P}\langle n \vert,
   \eqno(2.12) $$
independent of $\sigma$.  For each $p^\mu$, a little algebra gives
(for the diagonal)
$$ S_p(\sigma) =
   1 -\vert n \rangle_p {{}_p\langle} n \vert
   + {{h(p,\sigma-i\epsilon)}
   \over h(p,\sigma+i\epsilon)}
   \vert n \rangle_p {{}_p\langle} n \vert,
  \eqno(2.13) $$
where the last term contains a projection to the subspace ${\cal K}$.
\par Analyticity arguments give
$$ {{h(p,\sigma-i\epsilon)} \over h(,\sigma+i\epsilon)}={\sigma-
  \mu_p^*\over \sigma-\mu_p}e^{if_p(\sigma)}.
   \eqno(2.14)$$
If there is no singularity at $\infty$ (an exponentially bounded 
 singularity  would correspond, in the Lax-Phillips
terminology, to a ``trivial inner factor''), we obtain the simple form
$$ S_p(\sigma)= 1 -\vert n \rangle_p{{}_p\langle} n \vert+
   {\sigma- \mu_p^*\over \sigma-\mu_p}\vert n \rangle_p{{}_p
   \langle} n \vert 
   \eqno(2.15)$$
\par The resonant wave function is then given explicitly as a vector
in the Lax-Phillips Hilbert space as
$$ {}_{out}\langle \sigma p' \alpha \vert R \rangle_p = \delta^4(p-p')
2i(Im \mu_p) {\vert n \rangle_p )^\alpha \over \sigma -
\mu_p}. \eqno(2.16)$$
\par  There is no simpler non-trivial Lax-Phillips structure; the
Lee-Friedrichs model therefore provides the simplest example$^{26}$ of a
 Lax-Phillips scattering theory with a resonance.
\bigskip
\noindent
{\bf 3. Nonrelativistic limit}
\smallskip
\par The structure of the Galilean group requires that the
nonrelativistic limit contains elementary objects of definite
mass. The condition
$$ E - Mc^2 = \varepsilon < \infty \eqno(3.1)$$
as $c \rightarrow \infty$ has been found$^{27}$ to be an effective method for
taking this limit.  For example, for the free particle, the
Hamiltonian
$$ \eqalign{K_0 &= {p^\mu p_\mu \over 2M} = -{1 \over
Mc^2}(\varepsilon + Mc^2)^2 + {{\bf p}^2 \over 2M} \cr
&=  {{\bf p}^2 \over 2M} - \varepsilon + {\rm const} \cr} \eqno(3.2)$$
implies that 
$$ \eqalign{{dt \over d\tau } &= {\partial K_0 \over \partial E}\cr 
&= {\partial K_0 \over \partial  \varepsilon} = 1 \cr}, \eqno(3.3)$$
and hence in this limit $t=\tau$ (up to a constant).  If there are many
 particles, $K_0$
has the form of a sum over such terms, and all of the $\{t_i \}$ can
be put into correspondence with $\tau$, the Newtonian time.  In this
way, in refs. $(27)$, one shows that the relativistic micronanical
ensemble goes over to the nonrelativistic form, and quantum mechanical
wave packets, relativistically spread in the $t$ direction, contract
to have support in the neighborhood of $t \sim \tau$.
\par Examining the momentum conservation law for the $VN\theta$
vertex,  
$$ p_V^\mu = p_\theta^\mu + p_N^\mu,   \eqno(3.4)$$ 
which must be valid by translation invariance of the whole system, we
see that
$$ \varepsilon_V + M_Vc^2 = \varepsilon_\theta + M_\theta c^2 +
\varepsilon_N + M_N c^2.$$
Since $\varepsilon_V, \varepsilon_\theta, \varepsilon_N$ are bounded
as $c \rightarrow \infty$, it follows that 
$$ M_V = M_\theta + M_N,  \eqno(3.5)$$
the Galilean mass conservation law, and
$$ \varepsilon_V = \varepsilon_\theta + \varepsilon_N, \eqno(3.6)$$
a conservation law for the mass fluctuation residues in the
Galilean limit. 
\par  The relativistic mass inequality for a decaying
particle (the mass of the decaying system must be equal or greater
than the masses of the decay products), in the form required by the
Stueckelberg unperturbed Hamiltonian structure, can be written as
$$ \eqalign{-{p_V^\mu p_{V\mu} \over 2M_V}&= -{{\bf p}^2 \over 2M_V} +
{(\varepsilon + M_V c^2)^2 \over 2M_V c^2}
\cr &\geq -{({\bf p} - {\bf k})^2 \over 2M_N} + {(\varepsilon_N +
M_N c^2)^2 \over 2M_N c^2}\cr &- {{\bf k}^2 \over 2M_\theta} +
{(\varepsilon_\theta + M_\theta c^2)^2 \over 2 M_\theta c^2}. \cr}
 \eqno(3.7)$$  
The Galilean mass terms and linear $\varepsilon$ terms cancel, and
for $ (\varepsilon/c)^2 \rightarrow 0$,one obtains
$$ {({\bf p} - {\bf k})^2 \over 2M_V} + {{\bf k}^2 \over 2M_\theta} \geq 
{{\bf p}^2 \over 2 M_V}, \eqno(3.8)$$
a property that should be satisifed for a decaying particle (the
kinetic energy of the particles in the final state is greater or equal
to that of the initial particle). 
\par The spectral representation $\vert \sigma \alpha \rangle$ for the
problem is represented in the direct sum of the sectors $\{\langle
p_V,\beta \vert \sigma \alpha \rangle_0,  \langle p_N, p_\theta, \beta
\vert \sigma \alpha \rangle_0 \}$($p_V \equiv p, \ p_\theta \equiv
k$), solutions of 
$$ \bigl\{ -{{\bf p}^2 \over 2M_V} + {(\varepsilon_V + M_V c^2)^2
\over 2M_V c^2 } \bigr\} \langle
p_V,\beta \vert \sigma \alpha \rangle_0= \sigma \langle
p_V,\beta \vert \sigma \alpha \rangle_0, \eqno(3.9)$$
and 
$$\bigl\{ - {{\bf p}_N^2 \over 2M_N} + {(\varepsilon_N + M_N c^2)^2
\over 2M_N c^2}  - {{\bf k}^2 \over 2 M_\theta} + {(\varepsilon_\theta
+ M_\theta c^2)^2 \over 2 M_\theta c^2} \bigl\}\langle p_N, k, \beta
\vert \sigma \alpha \rangle_0= \sigma \langle p_N, k, \beta
\vert \sigma \alpha \rangle_0. \eqno(3.10)$$
Since $[E,t] = i \hbar$, it is also true that $[\varepsilon, t] = i
\hbar$, i.e., $\varepsilon \rightarrow i\partial_t$, as in the
nonrelativistic model of Flesia and Piron$^2$.  Note that in the limit
$c \rightarrow \infty$, the same (infinite) constant must be
subtracted from $\sigma$ in both equations.
  In the nonrelativistic model of Flesia and Piron$^2$, the structure
discussed by Howland$^{28}$, often utilized in dealing with time
dependent Hamiltonian theories, was used:
 $$ \eqalign{K_0 &= E + H_0 \cr
 K &= E + H  = K_0 + V \cr}
\eqno(3.11)$$
The model representation$^{29}$ is defined by$^5$
$$\eqalign{{{}_0\langle} s, \alpha \vert K_0 \vert t \beta \rangle_m
&= -i\partial_s {{}_0\langle} s, \alpha \vert t \beta
\rangle_m \cr &= \sum_{\beta '}(i \partial_t \delta_{\beta \beta'} +
H_0^{\beta' \beta} {{}_0\langle}s \alpha \vert t \beta' \rangle_m,
\cr}
\eqno(3.12)$$
where, generally, 
$$ {{}_0\langle} s, \alpha \vert K_o \vert f) = \sum_{\beta '}(-i
\partial_t \delta_{\beta \beta'} + H_0^{\beta \beta'} {{}_m\langle}t
\beta' \vert f). \eqno(3.13)$$
The relation $(3.12)$, in the form
$$-i(\partial_s + \partial_t) {{}_0\langle} s \alpha \vert t \beta
\rangle_m = \sum_{\beta'} H_0^{\beta' \beta} {{}_0\langle} s \alpha
\vert t \beta' \rangle_m,  \eqno(3.14)$$
is seen to determine the $s+t$ dependence
of the transformation function, but the $s-t$ dependence remains arbitrary.
Unitarity restricts its form; it can be shown that$^5$
$$ S^{\beta \beta'}(\sigma) = \sum_{\alpha'\alpha} U^{\alpha \beta
*}(\sigma) \bigl( S^{aux}\bigr)^{\alpha \alpha'} U ^{\alpha'
\beta'}(\sigma), $$
so that the Lax-Phillips $S$-matrix is related to the $S$-matrix of
the problem $H= H_0 + V$ in the auxiliary space  by a unitary
transformaiton with some similarity to dilation analytic
methods$^{30}$.
\par The non-relativistic limit of the relativistic results $(3.9)$
and $(3.10$ contain terms of order $1/c^2$ which remove the
arbitrariness of description of the transformation; its consequences
for the nonrelativistic problem are under investigation.
\bigskip
\noindent
{\bf 4. Conclusions}
\smallskip
 \par We have studied the application of Lax-Phillips quantum theory
 to a soluble relativistic quantum field theoretical model. In this
 model, we obtain the Lax-Phillips
 $S$-matrix explicitly as an inner function (the Lax-Phillips structure
 is defined pointwise on a foliation over the total energy-momentum of
 the system).  The structure of the Lee model $S$-matrix has a
 term with factorized numerator, corresponding to the transition
 matrix element of the interaction, and denominator $h(p,
 \sigma)$ which contains the zero inducing the resonance pole. The
 numerator factors can be identified as a vector field over the complex
 $\sigma$ plane. Foliating the $S$-matrix over the total energy
 momentum $p$, it takes on the form of a projection into the space
 complementary to the discrete subspace of the rank one potential of
 the model (for each point $\sigma,\, p$), plus an scalar inner
 function on the discrete subspace.   
 The vector field on the complex extension of the
 spectral representation (on the singular point, it corresponds to the
 projection into the resonant eigenstate), is independent
 of the spectral parameter up to a scalar multiplicative function. It
 then follows  that
 the projection is in fact independent of $\sigma$.
 This result leads to the conclusion that the properties of the $S$-matrix are
 essentially derived from the properties of a scalar inner function.   
\par This inner function consists of, in general, a rational factor,
 which contains all of the zeros and poles, and a singular factor
 (constructed with singular measure).  If the singular factor is
 exponentially bounded, it is, in the terminology of Lax and
 Phillips$^1$, a trivial inner function.  The application of this
 inner function does not change the resonance structure, but the
 functional form of the eigenfunctions and scattering states may be
 altered.
\par We then studied the rational case, the simplest possible model
 for a non-trivial Lax-Phillips theory, for which the inner function
 reduces to just the ratio $ (\sigma -  \mu_p^*) /(\sigma -
 \mu_p)$.  We therefore see, conversely, that the simplest model for a
 non-trivial Lax-Phillips theory corresponds to a rank one Lee
 model$^{26}$.
 
\bigskip
\noindent
{\it Acknowledgements\/}
\par We wish to thank S.L. Adler for suggesting the application of the
Lax-Phillips method to quantum field theory, and E. Eisenberg and
C. Piron for useful discussions. One of us (L.H.) wishes to thank C. Piron
 for his hospitality at the Department of Theoretical Physics, University 
of Geneva, for his hospitality during the final stages of this work.

\bigskip
\noindent
{\bf References}
\frenchspacing
\item{1.}P.D. Lax and R.S. Phillips, {\it Scattering Theory}, Academic
 Press, New York (1967).
\item{2.} C. Flesia and C. Piron, Helv. Phys. Acta {\bf 57}, 697
(1984).
\item{3.} L.P. Horwitz and C. Piron, Helv. Phys. Acta {\bf 66}, 694
(1993).
\item{4.} E. Eisenberg and L.P. Horwitz, in {\it Advances in Chemical
Physics}, vol. XCIX,  ed. I. Prigogine and S. Rice, Wiley, New York
(1997), p. 245.
\item{5.} Y. Strauss, L.P. Horwitz and E. Eisenberg, hep-th/9709036,
submitted for publication.
\item{6.} C. Piron, {\it Foundations of Quantum Physics},
Benjamin/Cummings, Reading (1976).
\item{7.} V.F. Weisskopf and E.P. Wigner, Z.f. Phys. {\bf 63}, 54
(1930); {\bf 65}, 18 (1930).
\item{8.} L.P. Horwitz, J.P. Marchand and J. LaVita,
Jour. Math. Phys. {\bf 12}, 2537 (1971); D. Williams,
Comm. Math. Phys. {\bf 21}, 314 (1971).
\item{9} L.P. Horwitz and J.-P. Marchand, Helv. Phys. Acta {\bf 42}
1039 (1969). 
\item{10.} L.P. Horwitz and J.-P. Marchand, Rocky Mountain Jour. of
Math. {\bf 1}, 225 (1971).
\item{11.} B. Winstein, {\it et al,\ Results from the Neutral Kaon
Program at Fermilab's Meson Center Beamline, 1985-1997\/},
FERMILAB-Pub-97/087-E, published on behalf of the E731, E773 and E799
Collaborations, Fermi National Accelerator Laboratory, P.O. Box 500,
Batavia, Illinois 60510. 
\item{12.} T.D. Lee, R. Oehme and C.N. Yang, Phys. Rev. {\bf 106} ,
340 (1957).
\item{13.} T.T. Wu and C.N. Yang, Phys. Rev. Lett. {\bf 13}, 380
(1964).
\item{14.} L.P. Horwitz and L. Mizrachi, Nuovo Cimento {\bf 21A}, 625
(1974); E. Cohen and L.P. Horwitz, hep-th/9808030; hep-ph/9811332,
 submitted for publication.
\item{15.}  W. Baumgartel, Math. Nachr. {\bf 69}, 107 (1975);
 L.P. Horwitz and I.M. Sigal, Helv. Phys. Acta
 {\bf 51}, 685 (1978); G. Parravicini, V. Gorini
 and E.C.G. Sudarshan, J. Math. Phys. {\bf 21}, 2208
 (1980); A. Bohm, {\it Quantum Mechanics: Foundations
 and Applications\/,} Springer, Berlin (1986); A. Bohm,  M. Gadella
and
 G.B. Mainland, Am. J. Phys. {\bf 57}, 1105 (1989); T. Bailey and
 W.C. Schieve, Nuovo Cimento {\bf 47A}, 231 (1978).
\item{16.} See, for example: 
S.L. Adler and L.P. Horwitz, {\it Structure and properties of
 Hughstons's stochastic extension of the Schr\"odinger equation},
 quant-ph/9909026, IASSNS-HEP-99/83, L.P Hughston,
 Proc. Roy. Soc. Lond. A {\bf 452}, 953 (1996) and references therein.
\item{17.} K.R. Parthasarathy, {\it An introduction to quantum
 stochastic calculus}, Monoraphs in Mathematics, Birkhauser, Basel
 (1992), and references therein, particularly to the work of
R.L. Hudson and K.R. Parthasarathy.
\item{18.} I.P. Cornfield, S.V. Formin and Ya. G. Sinai,
{\it Ergodic Theory}, Springer, Berlin (1982).
\item{19.} Y. Four\`es and I.E. Segal, Trans. Am. Math. Soc. {\bf 78},
385 (1955).
\item{20.} M. Rosenblum and J. Rovnyak, {\it Hardy Classes and
Operator Theory}, Oxford University Press, New York (1985).
\item{21.} For example, J.R. Taylor, {\it Scattering Theory},
 John Wiley and Sons,
 N.Y. (1972); R.J. Newton, {\it Scattering Theory of Particles
 and Waves}, McGraw Hill, N.Y. (1976).
\item{22.}  T.D. Lee, Phys. Rev. {\bf 95}, 1329 (1954);
K.O. Friedrichs, Comm. Pure and Appl. Math. {\bf 1}, 361 (1950);
 L.P. Horwitz, Found. of Phys. {\bf 25}, 39 (1995);  N. Shnerb 
and L.P. Horwitz, Phys. Rev. {\bf A48}, 4068
(1993); L.P. Horwitz and N. Shnerb, Found. of Phys. {\bf 28}, 1509 (1998).
  See
also, D. Cocolicchio, Phys. Rev. {\bf D57}, 7251 (1998).
\item{23.} E.C.G. Stueckelberg, Helv. Phys. Acta {\bf 14}, 322, 588
(1941); J. Schwinger, Phys. Rev. {\bf 82}, 664 (1951); R.P. Feynman,
        Rev. Mod. Phys. {\bf 20}, 367 (1948) and Phys. Rev. {\bf 80},
        440 (1950); L.P. Horwitz and C. Piron, Helv. Phys. Acta {\bf
        46}, 316 (1973); R. Fanchi, Phys. Rev {\bf D20},3108 (1979);
        A. Kyprianides, Phys. Rep. {\bf 155}, 1 (1986) (and references
 therein).
\item{24.} I. Antoniou, M. Gadella, I. Prigogine and P.P. Pronko,
        Jour. Math. Phys. {\bf 39}, 2995 (1998).
\item{25.} Y. Strauss and L.P. Horwitz, to be published, Found. of
Phys.
\item{26.} Conjecture of C. Piron, 1993, personal communication.
\item{27.} L.P. Horwitz and F. Rotbart, Phys. Rev. {\bf D24}, 2127
(1981); L.P. Horwitz, W.C. Schieve and C. Piron, Ann. Phys. {\bf 137},
        306(1981); L.P. Horwitz, S. Shashoua and W.C. Schieve, Physica
        {\bf A 161}, 300 (1989).
\item{28.} J.S. Howland, Springer Lecture Notes in Physics vol. 130,
        p. 163, Springer, Berlin (1979).
\item{29.} K.O. Friedrichs, Comm. Pure and Appl. Math. {\bf 1}, 361
(1948).
\item{30.} J. Aguilar and J.M. Combes, Comm. Math. Phys. {\bf 22} 280
(1971); J.M. Combes, Proc. Int. Cong. Math., Vancouver (1974);
        B. Simon, Ann. Math. {\bf 97}, 247 (1973).        
\vfill
\eject
\end
\bye